\documentclass[prl,aps,twocolumn,amsfonts,showpacs,superscriptaddress]{revtex4-1}

\usepackage{epsfig} \usepackage{psfrag} \usepackage{amsmath}
\usepackage{amssymb} \usepackage{color} \usepackage{graphicx}
\usepackage{subfigure}

\newcommand{\bc}{\begin{center}}
\newcommand{\ec}{\end{center}}
\def\ba#1{\begin{array}{#1}\displaystyle}
\newcommand{\ea}{\end{array}}

\newcommand{\beq}{\begin{equation}}
\newcommand{\eeq}{\end{equation}}
\newcommand{\beqa}{\begin{eqnarray}}
\newcommand{\eeqa}{\end{eqnarray}}

\newcommand{\bi}{\begin{itemize}}
\newcommand{\ei}{\end{itemize}}

\def\frc#1#2{\frac{#1}{#2}}

\newcommand{\p}{\partial}

\newcommand{\bra}{\langle}
\newcommand{\ket}{\rangle}

\begin{document}

\title{Far from equilibrium energy flow in quantum critical systems}
\author{M. J.  Bhaseen} \affiliation{Department of Physics, King's
  College London, Strand, London WC2R 2LS, United Kingdom}
\author{Benjamin Doyon} \affiliation{Department of Mathematics, King's
  College London, Strand, London WC2R 2LS, United Kingdom}
\author{Andrew Lucas} \affiliation{Department of Physics, Harvard
  University, Cambridge MA 02138, USA} \author{Koenraad Schalm}
\affiliation{Department of Physics, Harvard University, Cambridge MA
  02138, USA} \affiliation{Institute Lorentz, Leiden University, PO
  Box 9506, Leiden 2300 RA, The Netherlands}

\begin{abstract}
We investigate far from equilibrium energy transport in strongly
coupled quantum critical systems. Combining results from gauge-gravity
duality, relativistic hydrodynamics, and quantum field theory, we
argue that long-time energy transport occurs via a universal
steady-state for any spatial dimensionality. This is described by 
a boosted thermal state.
We determine the transport properties of this emergent 
steady state, including the average energy flow and its long-time
fluctuations.
\end{abstract} 

\date{\today}

\pacs{67.10.Jn, 11.25.Tq}

\maketitle

{\em Introduction.}--- In recent years there has been significant
interest in the behavior of strongly correlated systems out of
equilibrium. Experiments on cold atomic gases have raised questions
ranging from the nature of thermalization in one spatial dimension
($d=1$) \cite{Kinoshita:Newton,Rigol:Therm,Rigol:Breakdown} to the
dynamics of spontaneous symmetry breaking
\cite{Sadler:Spontaneous,Smith:Bosequench,Baumann:Dicke2}.
Theoretical attention has focused on the behavior of many body systems
following time-dependent protocols such as rapid quenches
\cite{Calabrese:Quench}.  A strong motivation is the
possibility of establishing universal results for the far from
equilibrium response; for a review see Ref.~\cite{Polkovnikov:RMP}.

Considerable insight into the dynamics of quantum systems has been
obtained using integrability and field theory techniques.  Recent work
in $d=1$ has established a universal regime of thermal transport when
two isolated critical systems are brought into instantaneous thermal
contact \cite{Doyon:Heat,Bernard:Noneq,Bernard:Time}. The predicted
steady state energy flow across the interface has been recently
observed \cite{Karrasch:Noneqtherm} using time-dependent Density
Matrix Renormalization Group (DMRG) at finite temperature
\cite{Karrasch:Finite,Karrasch:Reducing,Huang:Scaling}.  Extending
such results to higher dimensions is a major challenge and is the
motivation for this manuscript.  For recent experiments on thermal
transport and the thermal expansion of cold atomic gases see
Refs.~\cite{Brantut:Thermo,Schmidutz:QJT}.

We will use a combination of methods, including gauge-gravity duality
\cite{McGreevy:Notes,Hartnoll:Lectures}, relativistic hydrodynamics
and field theory techniques, to establish universal results for
non-equilibrium thermal transport in $d>1$. We will focus on
relativistic conformal field theories (CFTs) describing quantum
critical points with a linear dispersion $\epsilon = v|{\bf k}|$; we
will generically set $v=\hbar = k_{\mathrm{B}}=1$.  We argue that
thermal contact between strongly coupled quantum critical systems
gives rise to a universal homogeneous steady state with a
non-vanishing energy flow. Moreover, the energy transport in this far
from equilibrium steady state is fully described by a Lorentz boosted
thermal distribution. This governs not only the average energy current 
but also its fluctuations. 
\begin{figure}
   \includegraphics[width=3.2in,clip=true]{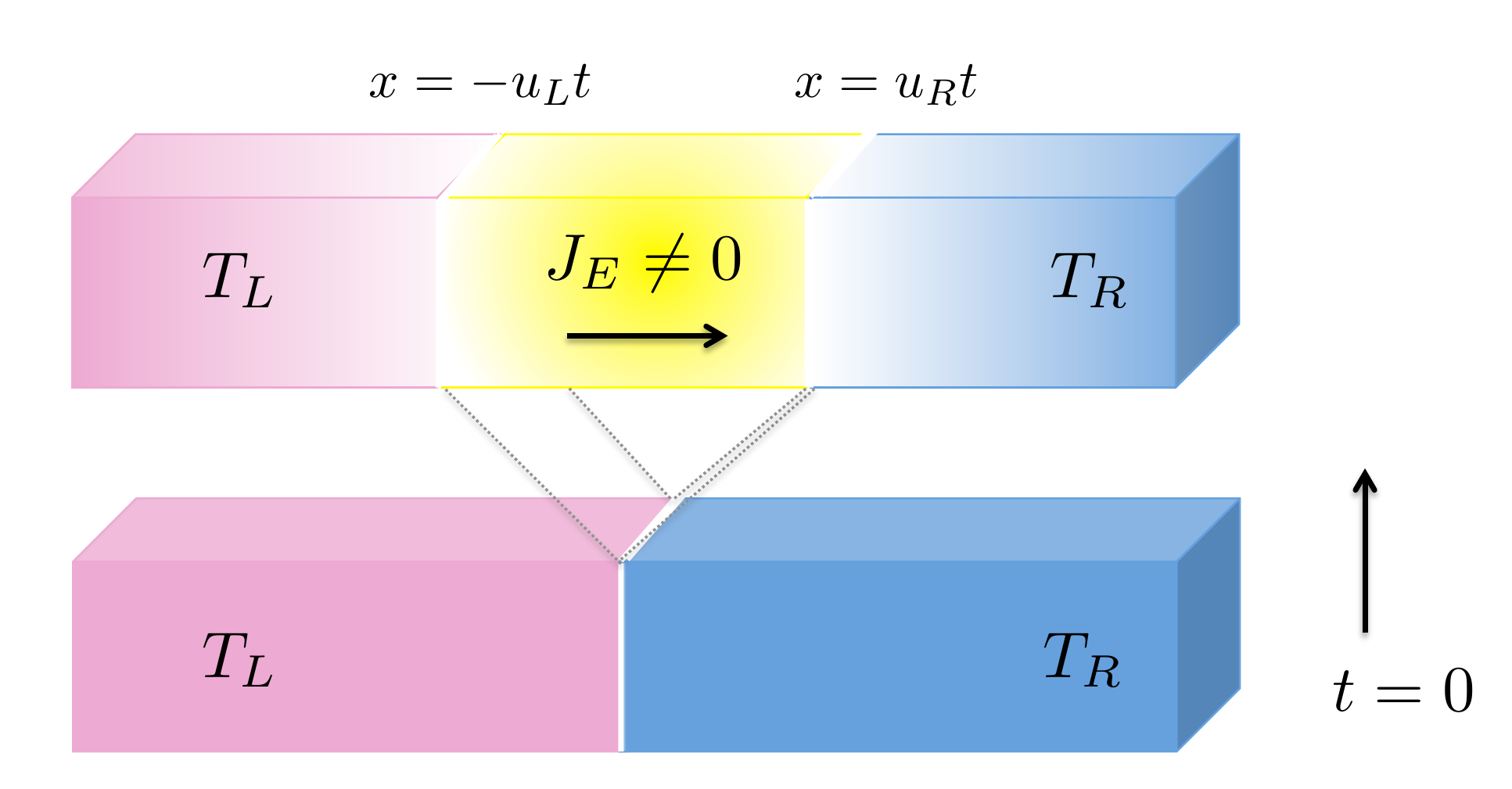} 
   \caption{Two isolated quantum critical systems at temperatures
     $T_{\rm L}$ and $T_{\rm R}$ are brought into instantaneous
     thermal contact. For large systems, and at late times, a
     spatially homogeneous non-equilibrium steady state develops
     across the interface. This carries an energy
     current $J_{{\rm E}}\equiv \langle T^{tx}\rangle_{\rm s}\neq 0$.}
   \label{Fig:Quench}
\end{figure}

{\em Setup.}--- We consider energy transport in a homogeneous quantum
critical system as depicted in Fig.~\ref{Fig:Quench}. The system is
subject to an initial temperature distribution with a step
profile. Equivalently, we bring into thermal contact the two
semi-infinite halves of the system that are independently
thermalized at left and right temperatures, $T_{\rm L}$ and $T_{\rm
  R}$.  A key question is whether a non-trivial current carrying
steady state emerges near the interface at late times; see
Fig.~\ref{Fig:Quench}. In particular, is there a steady state energy
flow with $J_{\rm E}\equiv\langle T^{tx}\rangle_{\rm s}\neq 0$? Here
$T^{\mu\nu}$ is the energy-momentum tensor and $s$ denotes the steady
state. If so, what is the value of this energy current and what are
the fluctuations in this steady state?

As we will argue below, a non-trivial current carrying state exists in
all dimensions.  This is in spite of the fact that there are no
external heat baths to drive a current. Rather, the semi-infinite
sub-systems themselves play the role of baths.  Although these
effective baths become asymptotically far apart at long times, the
steady state is expected to carry a current; at a quantum critical
point, the energy current is a conserved quantity and the transport
should have a ballistic component.

{\em One Dimension.}--- In $d=1$, the steady state of a CFT in the
above setup was shown to exist and is described in
Refs.~\cite{Doyon:Heat,Bernard:Noneq}.  In order to generalize these
results to higher dimensions it is instructive to examine these
findings using general field theory considerations.  First, in $d=1$,
one can show that the steady state is not accompanied by an energy
density gradient, but in fact the energy density must be homogeneous:
by conservation of $T^{\mu\nu}$ and stationarity, $\p_x \bra
T^{xx}\ket_{\rm s} = - \p_t \bra T^{tx}\ket_{\rm s}=0$.  Tracelessness
(scale invariance) yields $\bra T^{xx}\ket_{\rm s} = \bra
T^{tt}\ket_{\rm s}$ and so $\bra T^{tt}\ket_{\rm s}$ is
homogeneous. Second, conservation of $T^{\mu\nu}$ and tracelessness in
$d=1$ implies that the dynamics may be factorized into left- and
right-moving components.  With the initial condition of zero current,
this gives $\bra T^{tx}(x,t)\ket = F(x-t)-F(x+t)$ and $\bra
T^{tt}(x,t)\ket= F(x-t)+F(x+t)$, corresponding to sharp ``shock waves"
emanating from the interface at unit speed; see Fig.~\ref{Fig:Quench}.
Using the initial thermal form of the energy density on the left and
the right, $F(x) = (c\pi/12) T_{\rm L}^2 \Theta(-x) + (c\pi/12) T_{\rm
  R}^2\Theta(x)$, where $c$ is the central charge
\cite{Blote:Conformal,Affleck:Universal}. In the long time limit, the
steady-state energy current (for instance at $x=0$) is given by $\bra
T^{tx}\ket_{\rm s}= (c\pi/12)(T_{\rm L}^2-T_{\rm R}^2)$, corresponding
to the difference of independently ``thermalized" left- and
right-moving densities. Equivalently, $\bra T^{tx}\ket_{\rm
  s}=cg\Delta T$ where $\Delta T\equiv T_{\rm L}-T_{\rm R}$ and
$g=\pi^2k_B^2T_{\rm ave}/3h$ is the quantum of thermal conductance
\cite{Fazio:Anom, Rego:Quantized,Schwab:Measurement} with $T_{\rm
  ave}=(T_{\rm L}+T_{\rm R})/2$.

\begin{figure}
   \includegraphics[width=3.2in,clip=true]{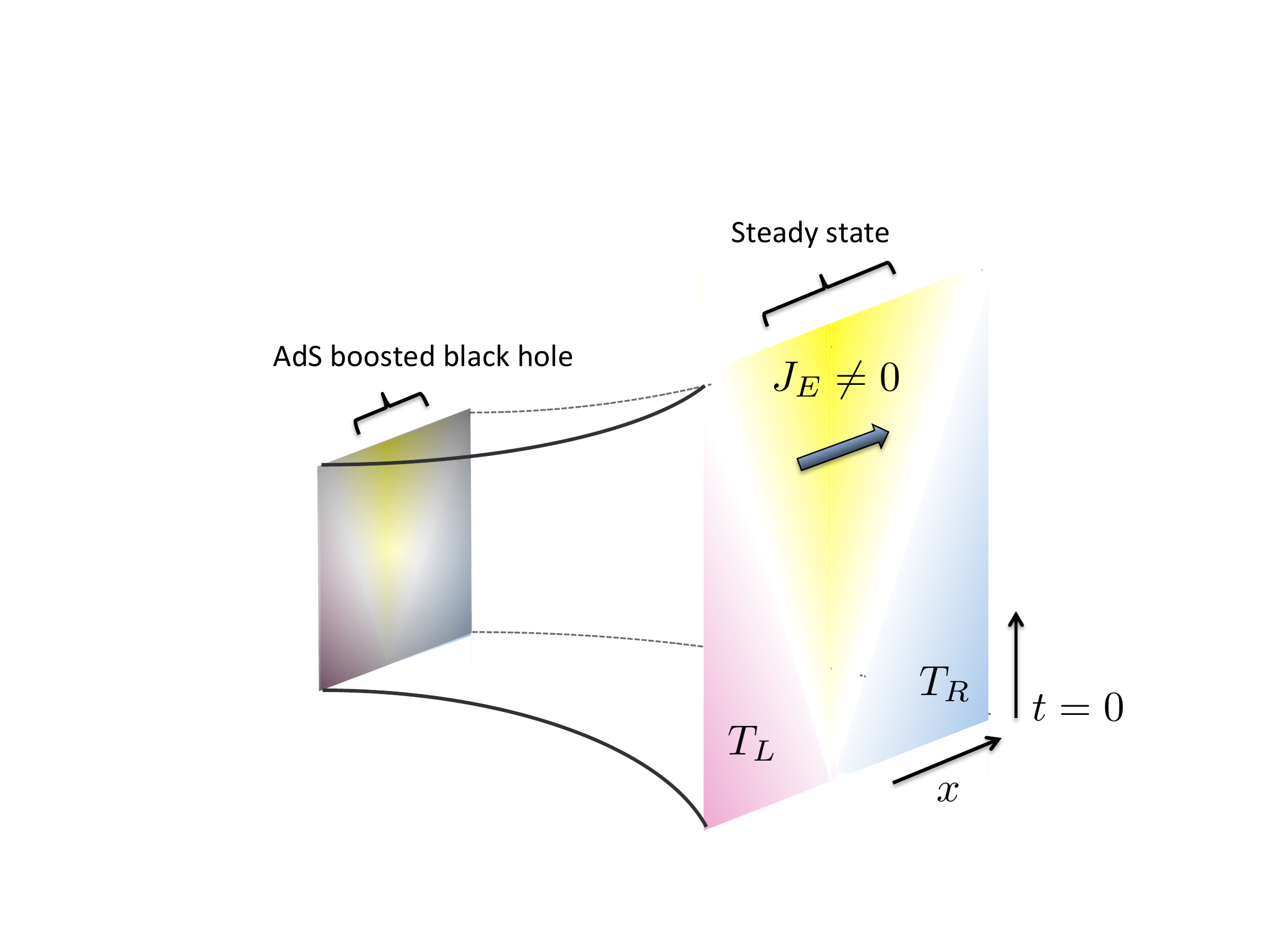} 
   \caption{Coordinates used for the AdS/CFT correspondence. The
     Hawking temperature of the unboosted black hole corresponds to
     the equilibrium temperature of the CFT. Lorentz boosted solutions
     describe far from equilibrium steady states. }
   \label{Fig:AdS}
\end{figure}

{\em Gauge-Gravity Duality.}--- What is the nature of the steady state
in higher dimensions? To answer this question, we first assume that
{\em in any dimension} a thermalization quench in a critical system
results in a completely homogeneous steady state with an energy flow;
we will provide {\em a posteriori} evidence for this later in the
manuscript.  In order to resolve the nature of this steady state we
employ gauge-gravity duality or holography
\cite{McGreevy:Notes,Hartnoll:Lectures}.  Gauge-gravity duality offers
unique opportunities for advancing our understanding of far from
equilibrium dynamics. The emergent behavior is encoded in the real
time evolution of black holes residing in AdS (Anti-de Sitter)
space-time, in one more spatial dimension; see Fig.~\ref{Fig:AdS}.
This has been used to explore thermalization in strongly coupled gauge
theories and the dynamics following quenches
\cite{Danielsson:Spherically,Bhattacharyya:Weak,Albash:Evolution,Das:Probe,
  Chesler:Horizon,Auzzi:Periodic,Murata:Noneq,Sonner:Hawking,Bhaseen:Holographic}. Although
technically the simplest form of the correspondence only holds for
strongly interacting theories with ``large-$N$ matrix" degrees of
freedom, holographic results can be viewed as a generalization of
Landau-Ginzburg theory that includes Wilsonian scaling. In particular
it allows access to the behavior of conformally invariant theories in
$d>1$, where there are very few tractable microscopic theories.

As we are concerned with the transport of energy, we study the
simplest holographic theory, which only contains Einstein--Hilbert
gravity:
\begin{equation}
S = \frac{1}{16\pi G_{\mathrm{N}}} \int \mathrm{d}^{d+2}x \; \sqrt{-g}(R-2\Lambda),
\label{holograv}
\end{equation}
where $\Lambda = -d(d+1)/2L^2$ is a negative cosmological constant and
$L$ is the radius of AdS.  The model (\ref{holograv}) is dual to a
strongly coupled CFT in $d$ spatial dimensions; see
Fig.~\ref{Fig:AdS}. In particular, the metric $g^{\mu\nu}$ is
dual to the energy-momentum tensor $T^{\mu\nu}$ of the CFT.  On the
gravitational side, $L$ should be large in units of Newton's constant
$L^d/G_{\mathrm{N}} \gg 1$, in order to use classical gravity. On the
dual gauge theory side, $L^d/G_{\mathrm{N}} \gg 1$ encodes the large
number of degrees of freedom of the CFT.

The homogeneous, stationary nature of the steady state should be
reflected in its gravitational dual. In a future publication
\cite{Future:Heatflow}, we will show that
{\em the only regular solutions to Einstein's equations, dual to theories on flat space-time, which encode
a homogeneous constant stress tensor
are the boosted black branes}: 
\begin{align}
  \mathrm{d}s^2 = \frac{L^2}{z^2}&\left[\frac{\mathrm{d}z^2}{f(z)} -
    f(z)(\mathrm{d}t\cosh\theta -\mathrm{d}x\sinh\theta)^2
    +\right.\nonumber\\ &\left. (\mathrm{d}{x}\cosh\theta -
    \mathrm{d}t\sinh\theta)^2+\mathrm{d}y_{\perp}^2\right], \label{ssmetric}
\end{align}
where
\begin{equation}
f(z) = 1-\left(\frac{z}{z_0}\right)^{d+1}\;\;\;\text{and}\;\;\; z_0 =
\frac{d+1}{4\pi T}.
\end{equation}
Here $\theta$ is the boost parameter corresponding to a boost in the
negative $x$-direction, $y_\perp$ parameterizes the transverse spatial
coordinates, $T$ is the {\em unboosted} temperature of the black hole,
and $z_0$ is the position of the planar horizon.
Hence, at least in the large-$N$ limit, the steady state is described
by a Lorentz boosted equilibrium state, after suitable identifications
of $T$ and $\theta$ in terms of $T_{{\rm L},{\rm R}}$.  We will argue that this 
also holds without large-$N$.

Following the rules of the AdS/CFT correspondence, $\langle
T^{\mu\nu}\rangle_{\rm s}$ is obtained from the metric of the dual
gravitational problem. After transformation to Fefferman-Graham
coordinates \cite{deHaro:Holo}:
\begin{align}
  \langle T_{\mu\nu}\rangle_{\rm s} &= \frac{L^{d}}{16\pi G_{\rm N}} \lim_{Z\rightarrow 0} \left(\frac{d}{dZ}\right)^{d+1} \frac{Z^2}{L^2}\,g_{\mu\nu}(z(Z)),
\end{align}
where $z(Z) =Z/R -(Z/R)^{d + 2}/[2(d + 1)z_0^{d + 1}]$ with $R=(d!)^{1/(d-1)}$. This yields the 
Lorentz boosted stress tensor of a finite temperature CFT 
\begin{equation}
   \langle T^{\mu\nu}\rangle_{\rm s} =
   a_d\,T^{d+1}\left(\eta^{\mu\nu}+(d+1)u^\mu u^\nu \right), 
\label{boostemtensor}
   \end{equation}
where $\eta^{\mu\nu} = \mathrm{diag}(-1,1,\cdots,1)$ is the CFT
metric and $u^{\mu}=(\cosh\theta,\sinh\theta,0,\ldots,0)$ is the resulting
velocity. The coefficient $a_d\sim L^{d}/G_{\rm N}$ characterizes the rest frame 
energy density of the CFT, $\langle T^{tt}\rangle=da_{d}\,T^{d+1}$ which is analogous to the Stefan--Boltzmann law \cite{Cardy:Ubiquitous}. In general $a_d$ is a measure of the number of degrees of freedom of the CFT, which 
depends on the details of the theory, including the strength of the coupling, see e.g. Ref.~\cite{Burgess:Free}.
The result (\ref{boostemtensor}) may also be obtained by direct Lorentz
transformation 
and the steady energy current is given by
\begin{equation}\label{currentboost}
   \langle T^{tx}\rangle_{\rm s} = \frc 12 a_d\,T^{d+1}(d+1)\sinh2\theta.
   \end{equation}
Here we recall that $\theta$ and $T$ are to be determined in terms of $T_{{\rm
    L},{\rm R}}$.  In $d=1$, the above picture is readily interpreted.
The boosted black hole corresponds to a state with its left- and
right-movers thermally populated at temperatures $T_{\rm
  L}=T\mathrm{e}^{\theta}$ and $T_{\rm R}=T\mathrm{e}^{-\theta}$
\cite{Kraus:Lectures}. Equivalently, $T_{\rm L}$ and $T_{\rm R}$ may
be regarded as the apparent temperatures arising from the Doppler
shift of the Stefan--Boltzmann radiation \footnote{We thank A. Green
  for suggesting this interpretation.}. Combining $a_1=L\pi/ 4
  G_{\mathrm{N}}$ with the relation $c= 3L/2G_{\mathrm{N}}$,
where $c$ is the central charge, one finds $a_1 = c \pi/6$. One
thus obtains the established non-equilibrium result $\langle
T^{tx}\rangle_{\rm s}=(c\pi/12)(T_{\rm L}^2-T_{\rm R}^2)$
\cite{Doyon:Heat}. 
For recent work examining energy flows dual to boosted black holes in different setups see Refs.~\cite{Marolf1,Figueras:Stationary,Marolf2}.

{\em Fluctuations.}--- A key observation is that the boosted state
(\ref{boostemtensor}) encodes not only the average energy current, but
also its fluctuations.  This may be illustrated in $d=1$, without
recourse to gauge-gravity duality and the large-$N$ limit. 
In $d=1$ the left- and right-movers are independently
thermalized, the exact steady state density matrix is given by
$\rho_{\rm s} = e^{-\beta_{\rm L}H_+ - \beta_{\rm R} H_-}$, where
$H_\pm=\sum_k(|k|\pm k)/2$ are the total energies of the right- and
left-moving excitations \cite{Doyon:Heat} and
$\beta_{{\rm L},{\rm R}}=1/T_{{\rm L},{\rm R}}$.  Equivalently, $H_\pm =
(H\pm P_x)/2$, where $H$ is the Hamiltonian and $P_x$ is the
total momentum. Therefore, $\rho_{\rm s} = e^{-\beta_+ H - \beta_-
  P_x}$ where $\beta_\pm = (\beta_{\rm L} \pm \beta_{\rm R})/2$.
 This is equivalent to
a boosted thermal state with $\rho=e^{-\beta\cosh \theta H + \beta
  \sinh\theta P_x}$, where $\beta=\sqrt{\beta_{\rm L}\beta_{\rm R}}$
is the inverse temperature in the rest-frame and $\theta$ is the boost
parameter, given by $e^{2\theta} = \beta_{\rm R}/\beta_{\rm L}$. The
non-equilibrium steady state in $d=1$
\cite{Doyon:Heat,Karrasch:Noneqtherm} is therefore also obtained by
``running past'' a thermal state with velocity $(\beta_{\rm
  R}-\beta_{\rm L})/(\beta_{\rm R}+\beta_{\rm L})$. Crucially, the
exact steady state density matrix $\rho_{\rm s}$ allows one to compute
not only the average energy flow, but also the exact generating
function of the energy current fluctuations \cite{Doyon:Heat,Bernard:Noneq,Bernard:Time}.

In general, one is interested in the full probability distribution of the integrated current density on the interface (that is, the total transfer of energy), $J = \int \mathrm{d}t \mathrm{d}y_\perp T^{tx}(x=0,y_\perp,t)$, in the steady state. Scaling out the large transverse area and the long time, the cumulants can be expressed in terms of connected correlation functions, $c_n\equiv\langle J^{n-1} T^{tx}(0)\rangle_{\mathrm{s}}^{\mathrm{c}}$. The generating function $F(z)\equiv\sum_{n=1}^\infty \frac{z^n}{n!} c_n$ is an important quantity, as non-equilibrium steady states are expected to give rise to nontrivial relations amongst cumulants encoded into the ``non-equilibrium fluctuation relations'' $F(\beta_{\mathrm{L}}-\beta_{\mathrm{R}}-z) = F(z)$ \cite{gallavotti1995,jarzynski2004,esposito2004}. Using AdS/CFT, cumulants are in principle computable from Witten diagrams holographically (see \cite{kovtun2005} for certain quadratic fluctuations). In [34], we will argue that, the steady state being a boosted thermal state, we can obtain the exact $F(z)$ in {\em any dimension}, whenever there is PT symmetry. Building on [11], we will argue that in this case the {\em extended fluctuation relations} (EFR) hold:
\begin{equation}\label{EFR}
	\frc{d F(z)}{dz} = J_E(\beta_{\mathrm{L}}-z,\beta_{\mathrm{R}}+z).
\end{equation}
Hence, the knowledge of the current as a function of the temperatures fixes $F(z)$. In particular, with parity symmetry $J_E(\beta_{\mathrm{L}},\beta_{\mathrm{R}}) = -J_E(\beta_{\mathrm{R}},\beta_{\mathrm{L}})$, the EFR implies the non-equilibrium fluctuation relations. In $d=1$ CFTs the EFR was shown to hold in [11]. In this case it is a direct consequence of left- and right-moving factorization.  This also allows us to derive this relation directly from thermal partition functions, equivalently from partition functions of the boosted black brane, $\mathrm{Tr} \; \mathrm{e}^{-\beta \cosh\theta\, H +\beta \sinh\theta\, P}$.

{\em Higher Dimensions.}--- For $d>1$ there is no holomorphic
factorization and thermalization is expected to modify the dynamics compared to $d=1$. We will show that it still develops a steady state. 
At long times, this should be
described by relativistic hydrodynamics. Indeed,
Eq.~(\ref{boostemtensor}) corresponds to the energy-momentum tensor of
a perfect conformal fluid, where $u^\mu$ is the local fluid velocity. 
The hydrodynamic equations simply express the conservation
of energy and momentum, $\partial_\mu \langle T^{\mu\nu}\rangle=0$. 
In a CFT we also have $\langle T_\mu^\mu\rangle=0$. Within this 
framework one may consider the effects of a range of
initial conditions that interpolate between asymptotic heat baths at
temperatures $T_{\rm L}$ and $T_{\rm R}$. At sufficiently large scales 
all of these will look like the initial conditions of the 
Riemann problem, $T_{\rm L}$ for $x<0$ and
$T_{\rm R}$ for $x>0$.

\begin{figure}
 \includegraphics[width=0.42\columnwidth,clip=true]{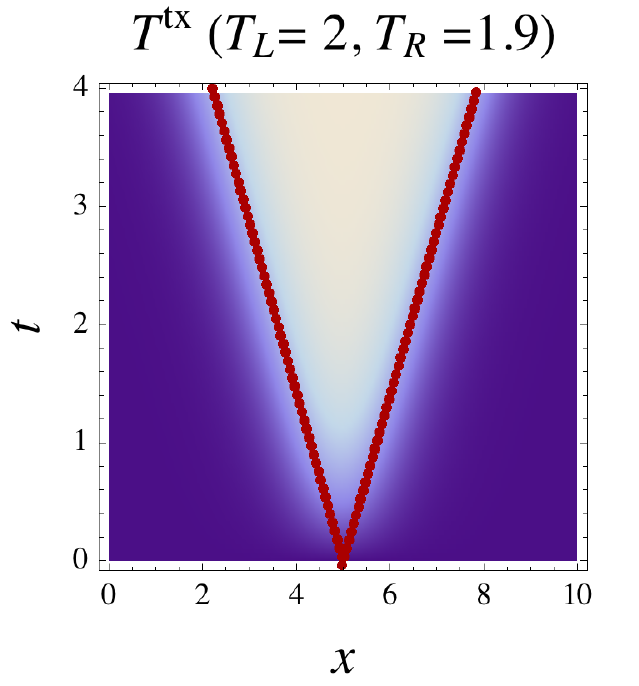}
 \includegraphics[width=0.42\columnwidth,clip=true]{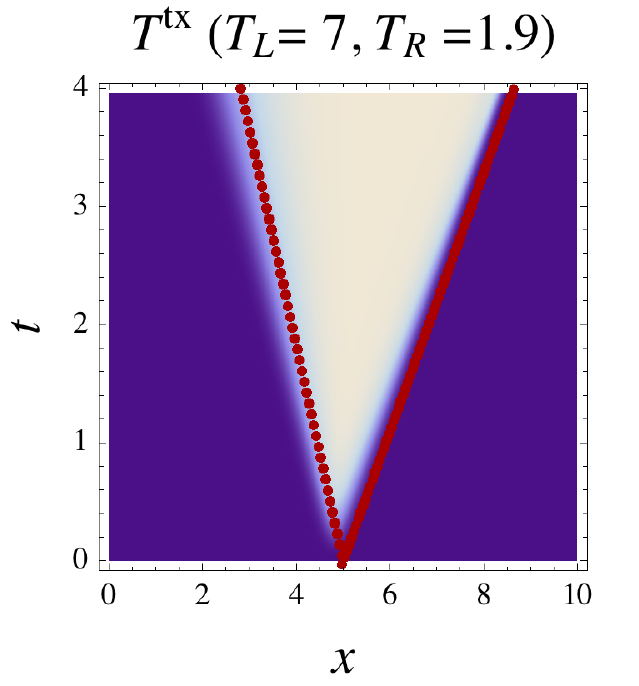}
 \includegraphics[width=0.13\columnwidth,clip=true]{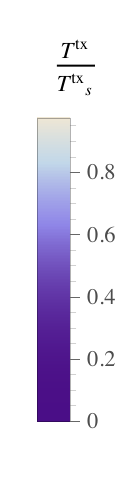}
 \includegraphics[width=0.49\columnwidth,clip=true]{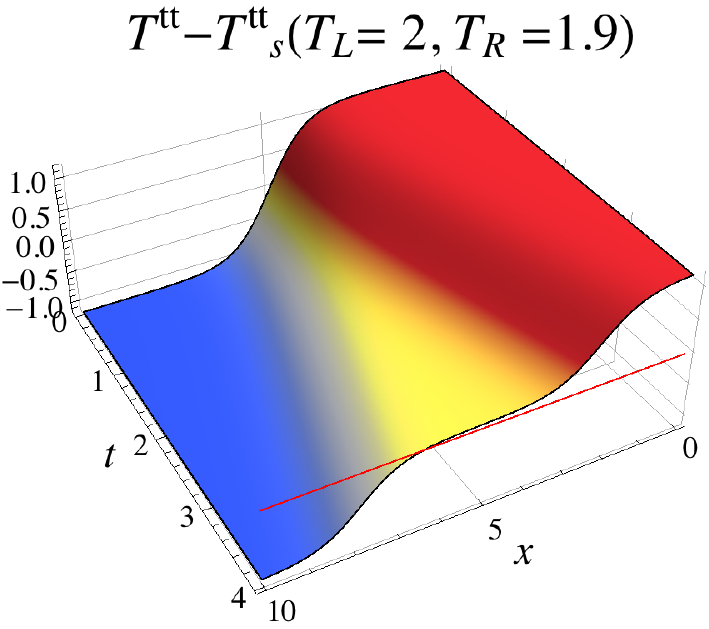}
 \includegraphics[width=0.49\columnwidth,clip=true]{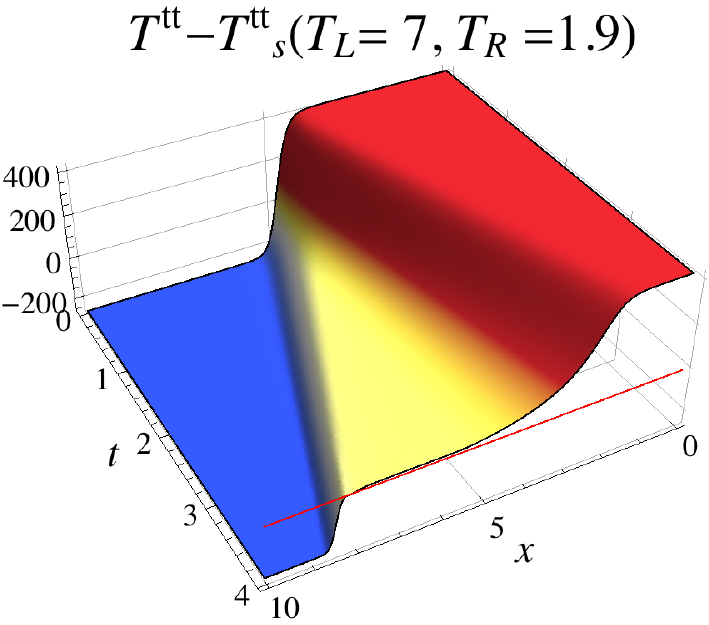}
\caption{Numerical solution of the relativistic continuity equation
     $\partial_\mu \langle T^{\mu\nu}\rangle=0$ with $\langle
     T_\mu^\mu\rangle=0$ in $d=2$. We take an initial temperature
     profile $T(x)=(T_{\rm L}+T_{\rm R})/2+[(T_{\rm R}-T_{\rm
         L})/2]\tanh(x/x_0)$.  (a) Density plot of $\langle
     T^{tx}\rangle$ showing outgoing shock
     waves with shock speeds $u_{{\rm L},{\rm R}}$ given by
     Eq.~(\ref{shockspeeds}). The analytical solutions are the red lines. 
For $T_{\rm L}\gg T_{\rm R}$ the asymmetry in the shock speeds is clearly visible. (b) Time-evolution of the energy density showing the approach to the
steady state solution. The red line is the predicted value of
$\langle T^{tt}\rangle_{\rm s}$ in the Lorentz boosted thermal state.}
   \label{Fig:Hydro}
\end{figure}

A solution consistent with these initial conditions consists of two
planar shock waves emanating from the contact region
\cite{Smoller}. We will now elucidate the properties of the steady
state in the intermediate region. We consider left- and right-moving
shocks that are homogeneous in the transverse spatial directions and
move at constant speeds $u_{\rm L}$ and $u_{\rm R}$ respectively; 
given the Riemann conditions the resulting solution is unique. Enforcing
energy and momentum conservation across the shocks constrains the form 
of the steady state energy-momentum tensor:
\begin{equation}
\langle T^{tx}\rangle_{\rm s} =a_d\,\left(
\frac{T_{\mathrm{L}}^{d+1}-T^{d+1}_{\mathrm{R}}}{u_{\mathrm{L}}+u_{\mathrm{R}}}\right),
\label{emconsshock}
\end{equation}
Invoking the boosted steady state (\ref{boostemtensor}) gives explicit expressions for $u_{{\rm L},{\rm R}}$ in terms of $T_{{\rm L},{\rm R}}$:
\begin{equation}
u_{\rm L}=\frac{1}{d}\sqrt{\frac{\chi+d}{\chi+d^{-1}}},\quad 
u_{\rm R}=\sqrt{\frac{\chi+d^{-1}}{\chi+d}},
\label{shockspeeds}
\end{equation}
where $\chi\equiv (T_{\rm L}/T_{\rm R})^{(d+1)/2}$. The result in the steady state 
region is a boosted thermal state, with 
temperature $T=\sqrt{T_{\rm L}T_{\rm R}}$ and boost velocity
given by $(\chi-1)/\sqrt{(\chi+d)(\chi+d^{-1})}$; in $d=1$ this reduces
to our previous result. The validity of these findings in $d>1$ 
is confirmed numerically in Fig.~\ref{Fig:Hydro}. 

The shock waves emanating from the contact region are non-linear
generalizations of sound waves. In particular, it follows from energy
and momentum conservation that the shock speeds satisfy the constraint
$u_{\rm L}u_{\rm R}=c_{\rm s}^2$, where $c_{\rm s}=1/\sqrt{d}$ is the
speed of sound. As a function of $T_{\rm L}/T_{\rm R}$, greater than unity,
 $u_{\rm R}$ interpolates between $c_s$ and $v$, whilst $u_{\rm L}$ interpolates 
between $c_s$ and $c_s^2/v$, where we have reinstated the microscopic velocity $v$.

A notable difference in $d>1$ compared to $d=1$ is that the system now
diffuses. By scaling, viscous corrections will not change the late
time results.  This is readily seen in the linear response regime,
$|T_{\mathrm{L}} - T_{\mathrm{R}}|\ll T_{\mathrm{L}} +
T_{\mathrm{R}}$, where we can solve the hydrodynamic equations
explicitly. One can show that the two ``shocks'' propagate at the
speed of sound, and have a width growing diffusively as $\sqrt{t}$. On
long length scales this reduces to the sharp shock dynamics discussed
above. In linear response the solution is stable, but large shear
perturbations may set off a turbulent instability. It would be
interesting to investigate this in future work.

{\em Conclusions.}--- We have established results for the far from
equilibrium energy flow in strongly coupled critical systems in
$d>1$. We predict the existence of steady state solutions with a
universal description for energy transport in terms of a boosted
thermal distribution.  Although we have focused on CFTs, we expect
that non-trivial steady states may 
emerge under a broader range of conditions, provided 
energy and momentum are conserved. It would be interesting to
verify these results in experiments using cold atomic gases, or in 
numerical simulations based on matrix product states.

{\em Acknowledgements.}--- We thank B. Benenowski, D. Bernard,
P. Chesler, A. Green, D. Haldane, C. Herzog, D. Marolf, B. Najian,
C.-A. Pillet, S. Sachdev and A. Starinets for helpful comments. MJB
and KS thank the Kavli Royal Society Center Chicheley Hall and the
Isaac Newton Institute, Cambridge for hospitality. MJB and BD thank
The Galileo Galilei Institute for Theoretical Physics. AL is supported
by the Smith Family Science and Engineering Graduate Fellowship. This
work was supported in part by a VICI grant of the Netherlands
Organization for Scientific Research (NWO), by the Netherlands
Organization for Scientific Reseach/Ministry of Science and Education
(NWO/OCW) and by the Foundation for Research into Fundamental Matter
(FOM).

\end{document}